\documentclass[12pt,preprint]{aastex}

\shorttitle{Multiwavelength Mapping of Coma}
\shortauthors{Eisenhardt et al.}

\begin{document}


\title{Multiaperture $UBVRIzJHK$ Photometry of Galaxies in the Coma Cluster }


\author{Peter R. Eisenhardt\altaffilmark{1,7},
        Roberto De Propris\altaffilmark{2},
        Anthony H. Gonzalez\altaffilmark{3},
        S. A. Stanford\altaffilmark{4,5},
        Mark Dickinson\altaffilmark{6},
        Michael C. Wang\altaffilmark{7,8}}

\altaffiltext{1}{Jet Propulsion Laboratory, California Institute of Technology,
MS169-327, 4800 Oak Grove Drive,
                 Pasadena, CA 91109}
\altaffiltext{2}{Cerro Tololo Inter-American Observatory, Casilla 603, La Serena,
                 Chile}
\altaffiltext{3}{Department of Astronomy, University of Florida, Gainesville,
                 FL 32611}
\altaffiltext{4}{Physics Department, University of California, Davis, 
                 1 Shields Avenue,  CA 95616}
\altaffiltext{5}{Institute of Geophysics and Planetary Physics, Lawrence
                 Livermore National Laboratories, Livermore, CA 94550}
\altaffiltext{6}{National Optical Astronomy Observatories, 950 N.
                 Cherry Avenue, Tucson AZ 85719}
\altaffiltext{7}{California Institute of Technology, 1200 E. California Boulevard,
Pasadena CA 91125}
\altaffiltext{8}{Kingbright Electronics, 3F, 317-1, Sec. 2, Chung Shan Rd.
Chung Ho, Taipei Hsien, Taiwan}

\begin{abstract}

We present a set of $UBVRIzJHK_s$ photometry for 745 $J+H$ band selected
objects in a $22.5' \times 29.2'$ region centered on the core of the Coma
cluster. This includes 516 galaxies and is at least 80\% complete to $H=16$, 
with a spectroscopically complete sample of 111 cluster members
(nearly all with morphological classification) for $H < 14.5$. For each object we
present total \cite{kron80} magnitudes and aperture photometry. As an 
example, we use these data to derive color-magnitude relations for Coma 
early-type galaxies, measure the intrinsic scatter of these relations and 
its dependence on galaxy mass, and address the issue of color gradients. 
We find that the color gradients are mild and that the intrinsic scatter about
the color-magnitude relation is small ($\sim 0.05$ mag in $U-V$ and less than 
$\sim 0.03$ in $B-R$, $V-I$ or $J-K$). There is no evidence that the intrinsic 
scatter varies with galaxy luminosity, suggesting that the cluster red sequence 
is established at early epochs over a range of $\sim 100$ in stellar mass.

\end{abstract}


\keywords{galaxies: formation and evolution; galaxies: photometry; galaxies: clusters: individual: Coma}


\section{Introduction}

Clusters of galaxies may provide the database needed for a coherent
theory of galaxy evolution, in the same way that clusters of stars meet
this need for stellar evolution.  Environmental effects such as tidal 
interactions and mergers, ram pressure stripping, and confinement of gas 
by the intra-cluster medium (to name a few), make galaxy clusters more 
complex than their stellar counterparts. However the great luminosities 
of galaxy clusters (and galaxies) compensate for such complications by 
making it possible to directly observe their evolution to large lookback 
times (for instance, Stanford, Eisenhardt \& Dickinson 1995, 1998, 
\citealt{depropris99,holden04,strazzullo06}) a luxury impractical at 
present for stars.

This hope has already been somewhat realized in a surprisingly
straightforward fashion, using the familiar color-magnitude (C-M)
diagram. A well defined ``main sequence" of luminous early-type galaxies 
is evident in nearby clusters and appears to have the same slope and
scatter in all systems (\citealt{sandage78}, Bower, Lucey \& Ellis 1992,
Terlevich, Caldwell \& Bower 2001, Lopez-Cruz, Barkhouse \& Yee 2004,
\citealt{mcintosh05}) and can be observed, essentially unchanged, at 
least to the highest redshifts studied \citep{stanford95,stanford98,
kajisawa00,vandokkum00,vandokkum01, blakeslee03,holden04,lidman04,wake05,
ellis06,holden06,mei06a,mei06b}.

The color-magnitude relation appears to be mainly due to a relation
between mass and metal abundance \citep{trager00,terlevich01}. Its
existence, and low scatter, may provide a stringent test of theories
of galaxy formation \citep{kaviraj05,renzini06}. The observations
imply a remarkably synchronized star formation history for early-type
galaxies across a wide range of environments, and straightforwardly modelled by
an initial burst of formation at high redshift followed by passive
evolution of the stellar population, similar to the early monolithic
collapse scenario of Eggen, Lynden-Bell \& Sandage (1962).

Measuring such changes with lookback time requires reference to
color-magnitude data at the same rest wavelengths at redshift zero.
The standard of reference used in \cite{stanford98}, as well
as many other similar studies, was the Coma cluster.  Rich in
early type galaxies and with a lookback time of only about 300 Myr,
multiband ({\it UBVRIzJHK}) observations of the Coma cluster were
compared to observations of clusters with large lookback times using
{\it interpolated} k-corrections. Selecting galaxies by their near-infrared
luminosity is desirable because this is representative of stellar mass (Gavazzi, 
Pierini \& Boselli 1996, Bell \& de Jong 2001) and is relatively insensitive
to dust and minor starbursts. The typical size of the fields observed by 
\cite{stanford95,stanford98}, and of the HST field of view at the redshifts 
of the clusters, is about 1 Mpc. For the Coma cluster, 1 Mpc corresponds to 
$\sim 30'$ (H$_0=67$). Hence obtaining and reducing the reference data, 
particularly in the near-infrared, was a challenging project. Because we 
expect that other workers will find an infrared-selected catalog of 
{\it UBVRIzJHK} Coma cluster photometry useful, we are publishing these 
data as a separate paper, with minimal analysis. Several studies in the 
literature (e.g., \citealt{shioya02,ellis06}) have used this database 
prior to publication and we believe that  this dataset will be useful 
for the general community. We have previously presented a study of the 
infrared luminosity function of Coma galaxies \citep{depropris98}.

The structure of this paper is as follows: section 2 presents our
observations and data reduction; section 3 presents the photometric
catalog; and section 4 provides an analysis of color-magnitude
relations and their intrinsic scatter, and a brief discussion of the
implications for galaxy formation models. All photometry is on the 
Vega system. For consistency with our previous work \citep{stanford95, 
stanford98}, we adopt a cosmology with $H_0=67$ km s$^{-1}$ Mpc$^{-1}$, 
$q_0=0.1$ and $\Lambda_0=0$. Adopting a redshift of 0.023, this gives a 
luminosity distance of 104 Mpc, and an angular scale of 0.482 arcseconds 
per kpc.  For the commonly used $H_0=70$ km s$^{-1}$ Mpc$^{-1}$, 
$\Omega_M=0.3$, $\Lambda_0=0.7$ cosmology, the corresponding values are 
100.2 and 0.464 respectively.

\section{Observations and Data Reduction}

\subsection{Infrared imaging}

Infrared imaging at $J$, $H$ and $K_s$ was obtained using the IRIM
camera, with a $256 \times 256$ NICMOS3 HgCdTe array at the KPNO 2.1~m
telescope on the night of 7 April 1993.  The pixels subtended
$\approx1.1''$, with the scale being slightly different in each band.
A $12 \times 9$ frame mosaic with $133''$ steps (i.e. $\approx53\%$
overlap) between frames in both axes was obtained in each band,
spanning 29.2'(RA) $\times 22.5'$(dec) centered on approximately
12:59:52.8 +27:55:00 (2000).  The specific area was chosen to include
the highest density region in Dressler's (1980) tabulation, including
Dressler's numbers 82, 91, and 168 (corresponding to  numbers 6, 26,
and 16 respectively in the present catalog) near the edges of the
region. Two frames were omitted from the mosaic: the H frame centered 
near 12:59:57.4 +28:01:39 had an anomalous sky level in one quadrant, 
and the K frame centered near 12:59:18.1 +27:57:13 was lost in the data
transfer process. 

The exposure time per frame was 15 seconds for $J$ and $K_s$ and 10 seconds 
for $H$. The mosaic corners were only sampled once, other positions along 
the edge were sampled twice, and positions more than $133''$ from the edge 
of the mosaic were sampled four times, resulting in a nominal total exposure 
time over most of the field of 60 s, 40 s and 60 s in $J$, $H$ and $K_s$ 
respectively. Figure 1 shows a $UVK$ mosaic of the entire survey region produced
from our data.

The images, which were taken using double correlated sampling, were
linearized using an empirical correction developed for each pixel,
and then flat fielded.  A $7 \times 7$ grid of observations of M67
was used to determine the type of flat field exposure which minimized
the photometric dispersion in measurements of the same stars at many
locations across the array.  For $J$ and $H$ median sky flats worked
best, while for $K_s$ an average of dome flats with ambient illumination
(i.e. lights off) minimized the dispersion, which was $\approx2\%$ for
all three bands. Next the DIMSUM\footnote {Deep Infrared Mosaicing
Software, developed by P. Eisenhardt, M. Dickinson, S.A. Stanford,
J. Ward, available at ftp://iraf.noao.edu/iraf/contrib/dimsum.tar.Z}
package within IRAF was used to carry out sky subtraction and masking
of bad pixels and cosmic ray hits. For each frame, DIMSUM calculates
a median sky frame from preceding and following frames, masking pixels
associated with detectable objects.  In the $J$ band a median of 10
surrounding frames while rejecting the 2 highest and 2 lowest values at
each pixel was found to produce the most uniform appearance across the
final mosaicked image.  For $H$ a median of 14 frames with rejection
of the 3 highest and 3 lowest pixels worked best.  For $K_s$ an 8 frame
median and rejecting 2 high and 2 low pixels was used for the top and bottom
third of the mosaic.  Because of the large extent of the two dominant
central galaxies (NGC 4874 and 4889), in the center three rows of the
mosaic DIMSUM tended to oversubtract the sky, and using a median which
rejected a larger number of frames substantially alleviated this problem.
Nevertheless the extended emission near these two galaxies has probably
been suppressed to some extent. 

Registration of each sky-subtracted frame to the nearest half pixel was
accomplished using offsets measured from objects which overlapped with
those in frames to the south or east. The resulting three IR mosaics were
registered to one another and rebinned to a common pixel size of $0.6845''$
using astrometry for 78 {\it HST} Guide Star Catalog objects within the
field. In the process the pixel scale for IRIM on the KPNO 2.1~m was
determined to be 1.0996, 1.0964, and 1.0922 $\pm 0.0005$ arcseconds per
pixel in $J$, $H$ and $K_s$ respectively.    

Multiple observations of five UKIRT standards transformed to the CIT
system \citep{elias82}, established that the IR data were
photometric, and that the extinction coefficients were 0.17, 0.07,
and 0.09 magnitudes per airmass in $J$, $H$ and $K_s$ respectively.
The Coma mosaic data were all obtained at airmass $< 1.15$.  Comparison
with the infrared photometry of Persson, Frogel \& Aaronson (1979)
(see below) showed no convincing evidence for a color term, but did reveal
the need for a correction of $2 - 4\%$ for light lost in the small
($5.5''$ diameter) apertures used in measuring the relatively faint UKIRT
standard stars. This aperture correction was determined empirically from
an average of over 50 brighter stars in the standard star images.
The zeropoints are judged to be accurate to $\pm 0.03$ magnitudes.

\subsection{Optical Data}

Optical data were obtained during service observations by George Jacoby
at the KPNO 0.9~m telescope on 15 and 16 March 1994 using a 2048$^2$ CCD
with $0.680''$ pixels. Exposures in $B$, $V$, $R$, $I$ were obtained in
two positions to cover the infrared mosaic, whereas three positions were
used for $U$ and $z$.  The total exposure times were 5400s in $U$, 1200s
in $B$, 500s in $V$ and $R$, 300s in $I$ and 1800s in $z$. Reduction was
carried out in the standard way for these images.  The $I$ and $z$ images
were obtained in photometric conditions and calibrated using observations
of \cite{landolt92} and \cite{thuan76} standards. The $z$ band transformations 
are judged to be accurate to only $\pm 0.15$ mag. Non--photometric data in $B$, 
$V$ and $R$ were recalibrated using observations of a $9.7' \times 9.7'$ portion 
of the Coma field and of 25 standards obtained in photometric conditions with the 
COSMIC instrument (using a 2048$^2$ CCD with $0.2846''$ pixels) on the Palomar 
200 inch telescope on 2 February 1995.  The $U$ band images from Kitt Peak 
were calibrated by matching photometry in large apertures on galaxies whose 
$U$ photometry was published by \cite{bower92}. The optical and infrared 
images were registered to a common coordinate system, degraded to the 
same resolution ($0.6845''$ per pixel) and blurred to the seeing of the 
worst image ($1.7''$ FWHM). The reduced and calibrated FITS files for 
the optical and infrared mosaics will be made publicly available through 
the NOAO Data Products Program.

\subsection{Photometry}

We used FOCAS \citep{jarvis81,valdes82} on the $J$ and $H$ images 
to produce two independent catalogs. The detection limit was chosen 
to be 3.5$\sigma$ above the sky level in an area equal to the PSF disk. 
The two catalogs were position matched to eliminate false detections.  
Only objects present in both catalogs were accepted in the final catalog 
(under the assumption that Coma galaxies have similar infrared colors so 
that the $J$ and $H$ images reach similar depths).

Astrometry for catalogued objects was determined using bright galaxies
to establish an initial solution using the IRAF task ccmap.  This
solution was then used to identify $\sim 50$ objects in the catalog
with $H < 15$ which were classified as stars by \cite{lobo97}. 
These stars were then used to determine the final astrometric 
solution, which had residual errors of 0.4 arcsec. Catalogued objects were 
identified with objects in existing catalogs if their positions agreed to 
within 3 arcseconds, using NED and \cite{lobo97}

The angular size of galaxies in the catalog varies widely, requiring the 
use of an adaptive aperture size for photometry.  Simple single aperture 
photometry would introduce excessive noise for faint objects if the aperture 
used is too large, or omit substantial light from larger galaxies in the 
opposite case. An additional problem is deciding which light should be 
associated with which galaxy, particularly in the central regions where 
galaxies are clearly overlapping.

Software designed for analysis of faint galaxies in moderately crowded
fields was kindly provided to us by Drs. L. Infante and C. J.  Pritchet
of the University of Victoria \citep{infante87}, containing provisions for
a (simple) excision of contaminating objects. For each object \cite{kron80}
image moments $r_1$ and $r_{-2}$ were computed, where $r_1$ is the first moment 
of the light profile and hence a measure of galaxy size, while $r_{-2}$ measures 
compactness and is defined by the deviation of the area of the galaxy under
the light profile from a point spread function.

Photometry in apertures of radius $2r_1$ is found to enclose most 
($ > 96\%$) of the total light \citep{infante87}. These apertures 
vary from object to object and may change from band to band.  From 
simulations with $r^{1/4}$ profiles, it was found necessary to ensure 
that the measurement aperture within which $r_1$ was determined had a 
radius at least $5 \times$ larger than $r_1$ to achieve a convergent value 
of $r_1$. For some of the brightest and largest objects this was impractical. 
For the brightest 10 galaxies 50 kpc (diameter) apertures were adopted to 
approximate total magnitudes.  Model light profiles were fit to the brightest two
galaxies (\# 1 and 3) and to the brightest star and the resulting
models subtracted before measuring photometry on the remaining objects.  
For the galaxies with $11 \lesssim H \lesssim 13$ fixed 30 kpc (diameter) 
apertures were used (for consistency with \citealt{stanford95,stanford98}).   
For fainter objects $2r_1$ apertures were used. These choices were found 
to be best in terms of photometric accuracy, stability, and noise. 

Star-galaxy separation was determined using the classification by \cite{lobo97}
which uses optical data taken in good seeing, and reaches $V=22.5$.  The only 
objects identified in the present catalog not found in other catalogs were numbers 
66, 225 and 711, all of which were outside the \cite{lobo97} area (which does 
not cover RA $ > $ 13:00:30 and Dec $<$  27:49:13 in the present survey area).  
Star-galaxy  separation in the region not surveyed by \cite{lobo97} was determined 
using the $r_{-2}$ compactness parameter \cite{kron80}. 

\section{Photometric catalog}

Table 1 presents our catalog of objects. In column 1 we show our ID
number, in order of decreasing $H$ luminosity (see discussion for Table
2 below). Equatorial coordinates (J2000) are given in columns 2 and 3.  
Column 4 shows our $H$ magnitude. Column 5 gives the classification as star or
galaxy, or if available its morphological type taken from \cite{dressler80,rood67,
caldwell93,graham03} or the classification shown in the Nasa Extragalactic Database
(NED) in a few cases (mostly for faint galaxies where the source of the NED classification
was unclear). The source for the morphological class is in column 6. Redshifts, 
compiled from the best values (lowest stated error) in NED, are in column 7. The 
sources for the redshifts are given in column 8. Identification numbers from the 
Godwin, Metcalfe \& Peach (1983) catalog are given in column 9. Cross identifications 
from the NGC, IC, \cite{rood67} and \cite{dressler80} catalogs are  shown in column 10.  
Notes on some objects are in column 11. We present the first lines of this table in the 
printed version, the remainder being available in electronic format.

Table 2 summarizes our estimates of total magnitude in all the bands.  The
layout of this table is repeated for all photometry tables to follow:
in column order we give our ID, $U$, $B$, $V$, $R$, $I$, $z$,
$J$, $H$ and $K$. Again, we show the first few lines of this table and
make the rest available electronically.

Tables 3-8 show the same information for apertures of $2.1'', $4.2'', $6.2''$,
$8.3''$, $10.3''$ (radius) and $r_1$. The aperture magnitudes (50 kpc and 30 kpc)
for the brightest objects are reported in Table 2. Finally, by way of example, 
we show total $H$ magnitudes  and $6.2''$ aperture colors for 8 colors of common 
astrophysical interest in Table 9 (first few lines in the printed version, 
with the rest of the table being made available electronically).

Magnitude errors were estimated via bootstrap simulations.  Representative 
galaxies over a six magnitude range in each band were replicated twenty 
times on a grid in the image, and their magnitudes extracted.  At a total
magnitude of $H \sim 15$ a signal to noise of 5 was achieved, consistent 
with expectations. Table 10 gives functional forms for the variations of the
error with magnitude in the $6\farcs2$ aperture, of the form $\sigma_{band}=
\exp(a*(mag)-b)$, where '`mag' is the magnitude in the filter `band'. 
Simulations in which galaxies with known magnitudes and 
$r_1$ were replicated and added to the data show that the catalog is at least 90\%
complete to H=16.5 for $r_1 \leq 2.05''$, and at least 80\% complete to H=16 for
$r_1 \leq 4.1''$.

\subsection{Comparison with previous photometry}

Table 11 compares our photometry to previous work. Where possible,
we derived magnitudes in exactly the same apertures as the comparison
measurements. The only exceptions are for \cite{doi95} and \cite{lobo97}, 
who quote isophotal magnitudes. For these we computed magnitudes in a circular 
aperture having radius equivalent to the semi-major axis of the limiting isophote.
Figure 2 shows the comparisons between the reference photometry and
our data.

Our photometry is generally within 2--3\% of the reference magnitudes.
Small differences in the central filter bandpass and calibration
errors can easily account for discrepancies at a few percent level.
However, we find a difference of 0.16 mags between our $U$ band data
(and those of \citealt{bower92}) and the photographic photometry of
\cite{strom78}). We also find a difference of 0.26 mags between
our work and \cite{strom78} in $R$, and of 0.21 magnitudes 
between this work and the Cousins $R$ data of \cite{secker97} and
\cite{jorgensen94}. Our photometry agrees with the Harris $R$ magnitudes of 
\cite{bernstein95}, taken with the same filter/detector combination, 
within 2\%.

We convolve the spectral energy distribution of a typical early-type
galaxy, with the sensitivity function for the detector used and the
filter bandpass, to derive predicted offsets in magnitude between our
data and the comparison work used. In the $R$ band, we find that the
predicted difference between our photometry and the work of \cite{jorgensen94}
and \cite{secker97} is $-0.172$ magnitudes, in good agreement with the $-0.210$
we measure. For the 127-04 emulsion and RG610 filter used by \cite{strom78}, we 
find a discrepancy of $-$0.26, identical to the observed value. Similarly, 
for the IIIa-J emulsion and UG-2 filter used by \cite{strom78} in the $U$, 
we predict a difference of 0.188 magnitudes, which compares well with the 
0.156 magnitudes found. Applying these corrections, our data lie within a 
few percent of all previous photometry. When we compare with \cite{lobo97} we also 
find a  difference of 0.08 magnitudes in the $V$ band: however, the different 
photometric methods employed (isophotal apertures vs. circular ones) are at 
least as important as differences in filter bandpasses and detectors. 

To facilitate comparisons to our photometry, we provide the response functions 
for our bands in tables 13 through 21, including the effects of filter and 
atmospheric transmission, and quantum efficiency.

\section{The color-magnitude relation of early-type galaxies in Coma}

As an example application of this dataset we study the color-magnitude
relation of early-type (E or S0) galaxies and its intrinsic scatter, for the 8
colors shown in Table 9, using the 111 galaxies brighter than $H=14.5$
where we have a complete sample of cluster members. Morphologies for
these galaxies are taken from the compilation of \cite{dressler80} or
\cite{rood67} in order of preference. Only 7 of these 111 galaxies
do not have a morphological classification from these sources and we 
accordingly do not use them for computation of color-magnitude relations
and scatter.

\cite{scodeggio01} has suggested that the actual slope of the color
magnitude relation is much flatter than measured here (and elsewhere)
and that this is due to the use of a fixed aperture for all galaxies
rather than an adaptive aperture based on the structural parameters of
each galaxy (e.g., using the half-light radius). Because galaxies have
internal color gradients, a fixed aperture samples more metal poor
populations for the fainter galaxies, thus steepening the color-magnitude
trend (which is largely a mass-metallicity trend -- \citealt{trager00,
terlevich01}).

To examine the effect of color gradients within galaxies on the derived
CMR's, we measured and fit the $U-V$ vs.\ $V$ CMR using a series of
fixed circular apertures with radii ranging from $2\farcs1$ to
$10\farcs3$.  We also used apertures  with radii $r_1$ and $2 \times
r_1$ where $r_1$ is the first moment of the light profile \citep{kron80}
and is calculated for each galaxy. The $r_1$ and $2 \times r_1$
apertures should not be affected by the $1.7''$ FWHM seeing, as they
are generally considerably larger. The intercept of the CMR is redder
for smaller apertures, as expected given the general sense of color
gradients that are observed within elliptical galaxies.  The slope,
however, changes only slightly with aperture size, from $-$0.128 for the
smallest aperture to $-$0.085 for $r = 2 \times r_1$.  By contrast 
\cite{scodeggio01} finds an essentially flat relation when measured 
within $r_e$ in $U-V$ vs $V$. However, our findings are consistent
with the mild color gradients observed for E/S0 galaxies in nearby clusters
by \cite{tamura01}.

Although it is important to remember that any derived CMR does depend on 
the apertures used for the photometry, particularly in its intercept, our 
basic results concerning the slope and scatter of the CMR are insensitive 
to the aperture size used, and we therefore restrict our attention to the 
fixed $6\farcs2$ radius apertures for the remainder of this paper. 

Figure 3 shows the color-magnitude relations and best linear regression
fits for the colors listed in Table 9, vs total $H$ band magnitude. 
(For reference,  $L^*$ in the $H$ band is 11.13 \citep{depropris98}).
In Figure 4 we show one of these color-magnitude relations using different
symbols for members and non-members (the plots in Figure 3 are too 
compressed to show these clearly).

The slope and intercepts of these fits are shown in Table 12. The intrinsic 
scatter, also tabulated in Table 12, is calculated using the bootstrap 
method of \cite{stanford95,stanford98}. One can see that the slope flattens 
for colors redder than $V$ and that the intrinsic scatter is approximately 
constant for colors which are equally spaced in (logarithmic) wavelength, 
suggesting that the relation is indeed driven primarily by metal abundance 
and that the majority of the stellar populations were formed at early epochs
(this is due to the fact that for old stellar populations the only age and
metallicity sensitive index lies in the region of the 4000 \AA\ break and
the Magnesium complex -- see \citealt{kodama97,vazdekis01}).

It is also interesting to calculate how the intrinsic scatter varies
with galaxy $H$ band luminosity (i.e., stellar mass): Figure 5 shows
the variation of the intrinsic scatter in 0.5 mag. bins for four
colors. We see no evidence that the intrinsic scatter varies with
galaxy luminosity (this may also be appreciated from Figure 3, where
the CMR does not appear to spread at faint luminosities).

The existence of a color-magnitude relation is ascribed to a relation between
galaxy mass and metallicity, which is suitably explained by models of
monolithic collapse followed by fast winds. The small scatter about the
relation implies that most of the stellar populations must be quite old
and that the timescale for galaxy formation is relatively short, while
few mergers may have occurred since early epochs \citep{bower92,bower99}.
Our observations confirm the small scatter seen by Bower, but also show that 
the small scatter extends to fainter ellipticals ($\sim L^*+3$). This suggests 
that the stellar populations of all cluster early-type galaxies, irrespective
of mass, may have formed rapidly and at high redshift, and that
the color-magnitude relation was formed at an early epoch. Indeed,
a mature relation is already observed in the $z=0.83$ cluster MS1054-0321
\citep{andreon06a}. A similar conclusion was also reached by \cite{andreon06b}
on the basis of their thin color-magnitude relation in Abell 1185. The fact 
that the intrinsic scatter of the CMR is small for both bright and faint
galaxies implies that all red sequence galaxies, irrespective of mass, have
undergone a rapid star formation history, in contrast with evidence
for `downsizing' among the faint field (and high redshift) galaxy population 
\citep{heavens04} or for a truncated sequence at higher redshifts \citep{delucia04}.

\acknowledgements
We thank George Jacoby for taking the optical data for us. We also thank
the anonymous referee for a very helpful report that helped make the paper
better. Portions of this research were carried out at Jet Propulsion
Laboratory, California Institute of Technology, under a contract with NASA.
SAS's work was performed under the auspices of the U.S. Department of  Energy,
National Nuclear Security Administration by the University of  California,
Lawrence Livermore National Laboratory under contract No. W-7405-Eng-48.


\begin{deluxetable}{lcccccccccc}
\tablewidth{0pt}
\tablecaption{Main catalog of Coma objects}
\rotate
\tablehead{
\colhead{ID} & \colhead{RA (2000)} & \colhead{Dec (2000)} &
\colhead{$H$} & \colhead{Type} & \colhead{Ref.} &\colhead{$cz$ (km/s)} 
& \colhead{Ref.} & \colhead{GMP83} & \colhead{NGC/IC/RB/D} & \colhead{Notes} }
\startdata
   1 &  13:00:08.06 & 27:58:37.4 &  9.14 &   Db  &  D80\tablenotemark{a}   &  6495. &  M02\tablenotemark{b} &  2921. & NGC4889,D148 & \\
   2 &  13:00:48.41 & 27:48:03.6 &  9.51 &  star &                     &     0. &                  &     0.  &  & \\                
   3 &  12:59:35.62 & 27:57:34.1 &  9.60 &   cD  &  D80                &  7224. &  S00             &   3329. & NGC4874,D129 & \\  
   4 &  12:58:56.78 & 27:51:31.4 & 10.23 &  star &                     &     0. &      &    0.      & & \\               
   5 &  12:59:28.85 & 27:56:14.4 & 10.23 &  star &                     &     0. &      &    0.      & & \\                
   6 &  13:00:55.99 & 27:47:26.6 & 10.25 &   Sb  &  D80                &  7985. &  H97 & 2374.      & NGC4911,D82 & \\         
   7 &  12:59:20.52 & 27:51:42.1 & 10.54 &  star &                     &     0. &     &     0.     & \\                
   8 &  13:00:54.33 & 28:00:28.1 & 10.55 &   E   &  D80                &  8793. & S00 & 2417.      & IC4051,D143 & \\         
   9 &  13:00:41.60 & 27:50:39.6 & 10.62 &  star &                     &     0. &     &    0.      &             & \\\             
  10 &  12:59:19.80 & 28:05:03.9 & 10.62 &   E   &  D80                &  4700. & S04 & 3561.      & NGC4865,D179 & \\       
\enddata
\tablenotetext{a}{Morphology references: D80 -- Dressler (1980); RB -- Rood \& Baum (1967);
C93 -- \cite{caldwell93}; GG -- \cite{graham03}}
\tablenotetext{b}{Redshift references; A00 -- \cite{adami00};
B95 -- \cite{biviano95}; C93 -- \cite{caldwell93}; C96 -- \cite{casoli96}; 
CD -- \cite{colless96}; C01 -- \cite{castander01}; D88 -- \cite{dressler88};
E02 -- \cite{edwards02}; H97 -- \cite{haynes97}; M99 -- \cite{mueller99}; 
MG -- \cite{matkovic05}; M01 -- \cite{mobasher01}; M02 -- \cite{moore02}; 
S00 - \cite{smith00};  S04 -- \cite{smith04}; }
\end{deluxetable}
\clearpage 
\begin{deluxetable}{lccccccccc}
\tablewidth{0pt}
\rotate
\tablecaption{Total magnitudes for Coma objects}
\tablehead{
\colhead{ID (Table 1)} & \colhead{$U$} & \colhead{$B$} & \colhead{$V$} 
& \colhead{$R$} & \colhead{$I$} & \colhead{$z$} & \colhead{$J$} & \colhead{$H$}
&\colhead{$K$}}
\startdata
1\tablenotemark{a} &  13.83 &  13.27 &  12.24 &  11.65 &  10.91 & 10.59 &  9.86 &   9.14 &  8.88 \\
2 &  13.88 &  13.64 &  12.79 &  12.65 &  11.56 & 11.54 & 10.12 &   9.51 &  9.76 \\
3\tablenotemark{b} &  14.24 &  13.74 &  12.71 &  12.12 &  11.39 & 11.03 & 10.37 &   9.60 &  9.41 \\
4 &  13.51 &  13.47 &  12.88 &  12.75 &  11.78 & 11.70 & 10.65 &  10.23 & 10.24 \\
5 &  14.68 &  14.09 &  13.20 &  12.91 &  11.87 & 11.82 & 10.61 &  10.23 & 10.20 \\
6 &  14.12 &  13.94 &  13.17 &  12.60 &  12.07 & 11.70 & 10.94 &  10.25 &  9.98 \\
7 &  16.23 &  14.97 &  13.91 &  13.42 &  12.34 & 12.19 & 11.14 &  10.54 & 10.45 \\
8 &  15.01 &  14.53 &  13.62 &  13.00 &  12.30 & 11.98 & 11.25 &  10.55 & 10.41 \\ 
9 &  13.20 &  13.57 &  12.93 &  12.87 &  11.97 & 11.96 & 10.91 &  10.62 & 10.63 \\
10 &  15.06 &  14.68 &  13.69 &  13.13 &  12.44 & 12.06 & 11.44 &  10.62 & 10.38 \\
\enddata
\tablenotetext{a}{The near-infrared magnitudes for these two galaxies are
likely to be overestimated because of their large extent, preventing accurate
sky subtraction -- see Section 2.1 for details}
\end{deluxetable}
\clearpage

\begin{deluxetable}{lccccccccc}
\tablewidth{0pt}
\rotate
\tablecaption{$2\farcs1$ Aperture magnitudes for Coma objects}
\tablehead{
\colhead{ID (Table 1)} & \colhead{$U$} & \colhead{$B$} & \colhead{$V$} 
& \colhead{$R$} & \colhead{$I$} & \colhead{$z$} & \colhead{$J$} & \colhead{$H$}
&\colhead{$K$}}
\startdata
1$^a$ & 15.94 & 15.33 & 13.69 & 13.10 & 12.87 & 12.12 & 11.79 & 11.03 & 10.77 \\
2 & 13.90 & 13.69 & 12.78 & 12.65 & 11.62 & 11.54 & 10.15 &  9.58 &  9.81 \\
3$^a$ & 16.66 & 16.09 & 14.38 & 13.80 & 13.67 & 12.78 & 12.58 & 11.81 & 11.57 \\
4 & 13.53 & 13.52 & 12.88 & 12.75 & 11.83 & 11.70 & 10.67 & 10.25 & 10.27 \\
5 & 14.70 & 14.10 & 13.20 & 12.91 & 11.91 & 11.82 & 10.63 & 10.26 & 10.23 \\
6 & 17.06 & 16.59 & 15.09 & 14.43 & 14.11 & 13.45 & 13.00 & 12.27 & 11.97 \\
7 & 16.24 & 15.02 & 13.91 & 13.41 & 12.38 & 12.19 & 11.16 & 10.57 & 10.48 \\
8 & 16.95 & 16.30 & 14.80 & 14.20 & 13.96 & 13.25 & 12.80 & 12.05 & 11.82 \\
9 & 13.22 & 13.62 & 12.93 & 12.87 & 12.02 & 11.96 & 10.93 & 10.65 & 10.67 \\
10 & 16.15 & 15.66 & 14.31 & 13.73 & 13.35 & 12.72 & 12.22 & 11.50 & 11.29 \\
\enddata
\tablenotetext{a}{See essential footnote in Table 1}
\end{deluxetable}
\clearpage

\begin{deluxetable}{lccccccccc}
\tablewidth{0pt}
\rotate
\tablecaption{$4\farcs2$ Aperture magnitudes for Coma objects}
\tablehead{
\colhead{ID (Table 1)} & \colhead{$U$} & \colhead{$B$} & \colhead{$V$} 
& \colhead{$R$} & \colhead{$I$} & \colhead{$z$} & \colhead{$J$} & \colhead{$H$}
&\colhead{$K$}}
\startdata
   1$^a$ & 15.94 & 15.33 & 13.69 & 13.10 & 12.87 & 12.12 & 11.79 & 11.03 & 10.77 \\
   2 & 13.90 & 13.69 & 12.78 & 12.65 & 11.62 & 11.54 & 10.15 &  9.55 &  9.81 \\
   3$^a$ & 16.66 & 16.09 & 14.38 & 13.80 & 13.67 & 12.78 & 12.58 & 11.81 & 11.57 \\
   4 & 13.53 & 13.52 & 12.88 & 12.75 & 11.83 & 11.70 & 10.67 & 10.25 & 10.27 \\
   5 & 14.70 & 14.12 & 13.20 & 12.91 & 11.91 & 11.82 & 10.63 & 10.26 & 10.23 \\
   6 & 17.06 & 16.59 & 15.09 & 14.43 & 14.11 & 13.45 & 13.00 & 12.27 & 11.97 \\
   7 & 16.24 & 15.02 & 13.91 & 13.41 & 12.38 & 12.19 & 11.16 & 10.57 & 10.48 \\
   8 & 16.95 & 16.30 & 14.80 & 14.20 & 13.96 & 13.25 & 12.80 & 12.05 & 11.82 \\
   9 & 13.22 & 13.62 & 12.93 & 12.87 & 12.02 & 11.96 & 10.93 & 10.65 & 10.67 \\
  10 & 16.15 & 15.66 & 14.31 & 13.73 & 13.35 & 12.72 & 12.22 & 11.50 & 11.29 \\
\enddata
\tablenotetext{a}{See essential footnote in Table 1}
\end{deluxetable}
\clearpage

\begin{deluxetable}{lccccccccc}
\tablewidth{0pt}
\rotate
\tablecaption{$6\farcs2$ Aperture magnitudes for Coma objects}
\tablehead{
\colhead{ID (Table 1)} & \colhead{$U$} & \colhead{$B$} & \colhead{$V$} 
& \colhead{$R$} & \colhead{$I$} & \colhead{$z$} & \colhead{$J$} & \colhead{$H$}
&\colhead{$K$}}
\startdata
   1$^a$ & 15.47 & 14.86 & 13.70 & 13.10 & 12.47 & 12.12 & 11.37 & 10.61 & 10.34 \\
   2 & 13.88 & 13.65 & 12.79 & 12.65 & 11.57 & 11.54 & 10.12 &  9.52 &  9.76 \\
   3$^a$ & 16.12 & 15.55 & 14.39 & 13.80 & 13.19 & 12.78 & 12.10 & 11.33 & 11.09 \\
   4 & 13.52 & 13.48 & 12.88 & 12.75 & 11.79 & 11.70 & 10.65 & 10.23 & 10.25 \\
   5 & 14.69 & 14.09 & 13.20 & 12.91 & 11.87 & 11.82 & 10.61 & 10.24 & 10.21 \\
   6 & 16.70 & 16.25 & 15.10 & 14.43 & 13.78 & 13.45 & 12.65 & 11.94 & 11.62 \\
   7 & 16.23 & 15.00 & 13.91 & 13.41 & 12.35 & 12.19 & 11.14 & 10.55 & 10.45 \\
   8 & 16.51 & 15.89 & 14.81 & 14.20 & 13.58 & 13.25 & 12.45 & 11.70 & 11.46 \\
   9 & 13.20 & 13.58 & 12.93 & 12.87 & 11.98 & 11.96 & 10.91 & 10.63 & 10.64 \\
  10 & 15.84 & 15.34 & 14.31 & 13.73 & 13.09 & 12.72 & 11.97 & 11.26 & 11.06 \\
\enddata
\tablenotetext{a}{See essential footnote in Table 1}
\end{deluxetable}
\clearpage

\begin{deluxetable}{lccccccccc}
\tablewidth{0pt}
\rotate
\tablecaption{$8\farcs3$ Aperture magnitudes for Coma objects}
\tablehead{
\colhead{ID (Table 1)} & \colhead{$U$} & \colhead{$B$} & \colhead{$V$} 
& \colhead{$R$} & \colhead{$I$} & \colhead{$z$} & \colhead{$J$} & \colhead{$H$}
&\colhead{$K$}}
\startdata
1$^a$ & 15.24 & 14.61 & 13.56 & 12.96 & 12.25 & 11.97 & 11.14 & 10.39 & 10.12 \\
2 & 13.87 & 13.64 & 12.78 & 12.63 & 11.55 & 11.52 & 10.11 &  9.51 &  9.75 \\
3$^a$ & 15.85 & 15.28 & 14.23 & 13.64 & 12.94 & 12.62 & 11.84 & 11.08 & 10.83 \\
4 & 13.51 & 13.46 & 12.87 & 12.74 & 11.77 & 11.69 & 10.65 & 10.22 & 10.24 \\
5 & 14.68 & 14.08 & 13.19 & 12.90 & 11.86 & 11.80 & 10.61 & 10.22 & 10.20 \\
6 & 16.39 & 15.97 & 14.94 & 14.27 & 13.55 & 13.30 & 12.41 & 11.70 & 11.37 \\
7 & 16.22 & 15.00 & 13.90 & 13.40 & 12.33 & 12.18 & 11.13 & 10.54 & 10.44 \\
8 & 16.26 & 15.66 & 14.68 & 14.07 & 13.37 & 13.12 & 12.24 & 11.50 & 11.26 \\
9 & 13.19 & 13.56 & 12.92 & 12.85 & 11.96 & 11.95 & 10.91 & 10.62 & 10.63 \\
10 & 15.68 & 15.20 & 14.22 & 13.64 & 12.95 & 12.63 & 11.85 & 11.15 & 10.95 \\
\enddata
\tablenotetext{a}{See essential footnote in Table 1}
\end{deluxetable}
\clearpage

\begin{deluxetable}{lccccccccc}
\tablewidth{0pt}
\rotate
\tablecaption{$10\farcs3$ Aperture magnitudes for Coma objects}
\tablehead{
\colhead{ID (Table 1)} & \colhead{$U$} & \colhead{$B$} & \colhead{$V$} 
& \colhead{$R$} & \colhead{$I$} & \colhead{$z$} & \colhead{$J$} & \colhead{$H$}
&\colhead{$K$}}
\startdata
1$^a$ & 15.09 & 14.48 & 13.43 & 12.83 & 12.12 & 11.85 & 11.01 & 10.25 & 9.99 \\
2 & 13.87 & 13.63 & 12.77 & 12.61 & 11.54 & 11.50 & 10.11 &  9.53 & 9.74 \\
3$^a$ & 15.70 & 15.13 & 14.08 & 13.50 & 12.79 & 12.46 & 11.70 & 10.93 & 10.69 \\
4 & 13.51 & 13.46 & 12.86 & 12.73 & 11.76 & 11.67 & 10.64 & 10.22 & 10.23 \\
5 & 14.68 & 14.08 & 13.19 & 12.89 & 11.85 & 11.79 & 10.60 & 10.22 & 10.19 \\
6 & 16.10 & 15.76 & 14.78 & 14.10 & 13.36 & 13.12 & 12.42 & 11.51 & 11.18 \\
7 & 16.22 & 14.99 & 13.90 & 13.39 & 12.33 & 12.17 & 11.13 & 10.53 & 10.44 \\
8 & 16.12 & 15.53 & 14.55 & 13.94 & 13.24 & 12.99 & 12.12 & 11.37 & 11.15 \\
9 & 13.19 & 13.55 & 12.91 & 12.84 & 11.96 & 11.93 & 10.90 & 10.61 & 10.62 \\
10 & 15.61 & 15.13 & 14.14 & 13.56 & 12.87 & 12.56 & 11.78 & 11.09 & 10.88 \\
\enddata
\tablenotetext{a}{See essential footnote in Table 1}
\end{deluxetable}
\clearpage

\begin{deluxetable}{lccccccccc}
\tablewidth{0pt}
\rotate
\tablecaption{$r_1$ Aperture magnitudes for Coma objects}
\tablehead{
\colhead{ID (Table 1)} & \colhead{$U$} & \colhead{$B$} & \colhead{$V$} 
& \colhead{$R$} & \colhead{$I$} & \colhead{$z$} & \colhead{$J$} & \colhead{$H$}
&\colhead{$K$}}
\startdata
1$^a$ & 15.09 & 14.48 & 13.43 & 12.83 & 12.12 & 11.85 & 11.01 & 10.25 &  9.99 \\
2 & 13.87 & 13.63 & 12.77 & 12.61 & 11.54 & 11.50 & 10.11 &  9.53 &  9.74 \\
3$^a$ & 15.70 & 15.13 & 14.08 & 13.50 & 12.79 & 12.46 & 11.70 & 10.93 & 10.69 \\
4 & 13.51 & 13.46 & 12.86 & 12.73 & 11.76 & 11.67 & 10.64 & 10.22 & 10.23 \\
5 & 14.68 & 14.08 & 13.19 & 12.89 & 11.85 & 11.79 & 10.60 & 10.22 & 10.19 \\
6 & 16.10 & 15.76 & 14.78 & 14.10 & 13.36 & 13.12 & 12.24 & 11.51 & 11.18 \\
7 & 16.22 & 14.99 & 13.90 & 13.39 & 12.33 & 12.17 & 11.13 & 10.53 & 10.44 \\
8 & 16.12 & 15.53 & 14.55 & 13.94 & 13.24 & 12.99 & 12.12 & 11.37 & 11.15 \\
9 & 13.19 & 13.55 & 12.91 & 12.84 & 11.96 & 11.93 & 10.90 & 10.61 & 10.62 \\
10 & 15.61 & 15.13 & 14.14 & 13.56 & 12.87 & 12.56 & 11.78 & 11.09 & 10.88 \\
\enddata
\tablenotetext{a}{See essential footnote in Table 1}
\end{deluxetable}
\clearpage

\begin{deluxetable}{lccccccccc}
\tablewidth{0pt}
\rotate
\tablecaption{Selected colors of Coma galaxies}
\tablehead{
\colhead{ID (Table 1)} & \colhead{$H$} & \colhead{$U-B$} & \colhead{$U-V$} &
\colhead{$B-V$} & \colhead{$B-R$} & \colhead{$V-I$} & \colhead{$V-K$} &
\colhead{$I-K$} & \colhead{$J-K$}}
\startdata
   1 &  9.14 &  0.66 &  1.83 &  1.17 &  1.76 &  1.28 &  3.41 & 2.13 & 1.03  \cr
   3 &  9.60 &  0.62 &  1.78 &  1.17 &  1.75 &  1.25 &  3.35 & 2.10 & 1.01  \cr
   6 & 10.25 &  0.50 &  1.66 &  1.16 &  1.83 &  1.36 &  3.53 & 2.16 & 1.03  \cr
   8 & 10.55 &  0.67 &  1.75 &  1.08 &  1.68 &  1.28 &  3.40 & 2.12 & 0.99  \cr
  10 & 10.62 &  0.55 &  1.59 &  1.03 &  1.61 &  1.27 &  3.30 & 2.03 & 0.91  \cr
  12 & 10.76 &  0.55 &  1.61 &  1.06 &  1.66 &  1.30 &  3.37 & 2.07 & 0.96  \cr
  14 & 10.87 &  0.56 &  1.65 &  1.09 &  1.68 &  1.28 &  3.28 & 2.00 & 0.89  \cr
  15 & 10.88 &  0.51 &  1.61 &  1.10 &  1.67 &  1.23 &  3.28 & 2.05 & 0.97  \cr
  16 & 10.94 &  0.57 &  1.61 &  1.04 &  1.65 &  1.30 &  3.43 & 2.13 & 0.94  \cr
  17 & 10.98 &  0.48 &  1.55 &  1.07 &  1.64 &  1.25 &  3.28 & 2.03 & 0.99  \cr
  20 & 11.26 &  0.53 &  1.56 &  1.02 &  1.63 &  1.32 &  3.40 & 2.08 & 1.01  \cr
  21 & 11.26 &  0.56 &  1.68 &  1.11 &  1.67 &  1.26 &  3.33 & 2.07 & 0.99  \cr
  23 & 11.30 &  0.54 &  1.57 &  1.03 &  1.60 &  1.29 &  3.33 & 2.04 & 0.99  \cr
  24 & 11.32 &  0.55 &  1.64 &  1.09 &  1.67 &  1.27 &  3.32 & 2.05 & 1.00  \cr
  25 & 11.37 &  0.54 &  1.65 &  1.11 &  1.69 &  1.27 &  3.34 & 2.07 & 1.00  \cr
  26 & 11.39 &  0.52 &  1.58 &  1.06 &  1.60 &  1.24 &  3.22 & 1.98 & 0.98  \cr
  27 & 11.40 &  0.54 &  1.59 &  1.05 &  1.63 &  1.27 &  3.28 & 2.01 & 0.96  \cr
  28 & 11.42 &  0.47 &  1.53 &  1.07 &  1.63 &  1.24 &  3.18 & 1.94 & 0.95  \cr
  29 & 11.44 &  0.51 &  1.58 &  1.07 &  1.64 &  1.24 &  3.25 & 2.01 & 0.94  \cr
  30 & 11.44 &  0.45 &  1.49 &  1.05 &  1.61 &  1.21 &  3.14 & 1.93 & 0.96  \cr
\enddata
\end{deluxetable}
\clearpage

\begin{deluxetable}{ccc}
\tablewidth{0pt}
\tablecaption{Parameters for errors in $6.2''$ aperture}
\tablehead{\colhead{Band} & \colhead{$a$} & \colhead{$b$}}
\startdata
U & 0.582 & 13.94 \\
B & 0.620 & 15.02 \\
V & 0.715 & 16.25 \\
R & 0.647 & 14.42 \\
I & 0.781 & 15.73 \\
z & 0.715 & 14.61 \\
J & 0.909 & 16.40 \\
H & 0.776 & 13.65 \\
K & 0.805 & 13.81 \\
\enddata
\end{deluxetable}
\clearpage

\begin{deluxetable}{lccc}
\tablewidth{0pt}
\tablecaption{Comparison with previous work}
\tablehead{\colhead{Band} & \colhead{$\Delta$mag (reference $-$ this work)} &
\colhead{$\sigma$} & \colhead{Reference} }
\startdata
  U & 0.028 & 0.037 & \cite{bower92} -- BLE92 \\
    & 0.156 & 0.059 & \cite{strom78} \\
  B & 0.028 & 0.047 & \cite{strom78}\\
    & 0.016 & 0.107 & \cite{doi95} \\
  V & 0.024 & 0.020 & \cite{bower92}\\
    & 0.019 & 0.042 & \cite{strom78} \\
    & 0.078 & 0.104 & \cite{lobo97}\\
  R & --0.261 & 0.108 & \cite{strom78} \\
    & --0.020 & 0.088 & \cite{bernstein95} \\
    & --0.210 & 0.134 & \cite{jorgensen94} \\
    & --0.213 & 0.023 & \cite{secker97} \\
  J & --0.019 & 0.049 & \cite{bower92} \\
    & --0.021 & 0.046 & \cite{persson79} -- PFA79 \\
  H & 0.010   & 0.051 & \cite{bower92} \\
    & --0.002 & 0.047 & \cite{persson79} \\
  K & --0.012 & 0.041 & \cite{bower92} \\
    & --0.009 & 0.049 & \cite{persson79} \\
\enddata
\end{deluxetable}
\clearpage

\begin{deluxetable}{lcccccc}
\tablewidth{0pt}
\rotate
\tablecaption{Color-magnitude relations and scatter}
\tablehead{\colhead{Color} & \colhead{Slope} & \colhead{Intercept} & \colhead{Scatter (measured)} &
\colhead{Photometric error} & \colhead{Intrinsic scatter} }
\startdata
$U-B$ & $-0.079 \pm 0.007$ & $1.386 \pm 0.089$ & $0.041 \pm 0.005$ & $0.016 \pm 0.002$ & $0.038 \pm 0.006$ \\
$U-V$ & $-0.122 \pm 0.010$ & $2.929 \pm 0.123$ & $0.050 \pm 0.004$ & $0.016 \pm 0.002$ & $0.047 \pm 0.004$ \\
$B-V$ & $-0.042 \pm 0.004$ & $1.543 \pm 0.049$ & $0.028 \pm 0.004$ & $0.008 \pm 0.001$  & $0.027 \pm 0.004$ \\
$B-R$ & $-0.055 \pm 0.006$ & $2.259 \pm 0.074$ & $0.030 \pm 0.004$ & $0.008 \pm 0.001$  & $0.029 \pm 0.004$ \\
$V-I$ & $-0.029 \pm 0.005$ & $1.540 \pm 0.066$ &  $0.033 \pm 0.003$ & $0.010 \pm 0.001$  & $0.031 \pm 0.003$ \\
$V-K$ & $-0.109 \pm 0.017$ & $4.455 \pm 0.209$ & $0.074 \pm 0.010$ & $0.015 \pm 0.002$  & $0.072 \pm 0.010$ \\
$I-K$ & $-0.080 \pm 0.013$ & $2.915 \pm 0.164$ & $0.066 \pm 0.009$ & $0.012 \pm 0.002$  & $0.065 \pm 0.009$ \\
$J-K$ & $-0.025 \pm 0.009$ & $1.232 \pm 0.111$ & $0.036 \pm 0.006$ & $0.016 \pm 0.002$  & $0.025 \pm 0.007$ \\
\enddata
\end{deluxetable}
\clearpage

\begin{deluxetable}{lc}
\tablewidth{0pt}
\tablecaption{$U$ band response function}
\tablehead{\colhead{Wavelength \AA} & \colhead{Throughput \%} }
\startdata
3050. &  0            \\
3100. &  0.000117987  \\
3150. &  0.0010573    \\
3200. &  0.00422111   \\
3250. &  0.0142508    \\
3350. &  0.0773872    \\
3400. &  0.137695     \\
3450. &  0.223221     \\
3500. &  0.334892     \\
3550. &  0.475813     \\
\enddata
\end{deluxetable}
\clearpage

\begin{deluxetable}{lc}
\tablewidth{0pt}
\tablecaption{$B$ band response function}
\tablehead{\colhead{Wavelength \AA} & \colhead{Throughput \%} }
\startdata
3600.  &  0 \\
3610.  &  0 \\
3620.  & 0.00419808 \\
3630.  & 0.0147285 \\
3640.  & 0.021926  \\
3650.  & 0.0302643 \\
3660.  & 0.0444033 \\
3670.  & 0.0587523 \\
3680.  & 0.0736034 \\
3690.  & 0.0889334 \\
\enddata
\end{deluxetable}
\clearpage

\begin{deluxetable}{lcccc}
\tablewidth{0pt}
\tablecaption{$V$ band response function}
\tablehead{\colhead{Wavelength \AA} & \colhead{Throughput} }
\startdata
4770. &  0 \\
4780. &  0.000172793 \\
4790. &  0.000690248 \\
4800. &  0.0117209 \\
4810. &  0.0175596 \\
4820. &  0.0259633 \\
4830. &  0.0372646 \\
4840. &  0.0521359 \\
4850. &  0.0710756 \\
4860. &  0.0944066 \\
\enddata
\end{deluxetable}
\clearpage

\begin{deluxetable}{lc}
\tablewidth{0pt}
\tablecaption{$R$ band response function}
\tablehead{\colhead{Wavelength \AA} & \colhead{Throughput} }
\startdata
5400. &  0.00703248 \\
5410. &  0.0112412 \\
5420. &  0.014039 \\
5430. &  0.0281244 \\
5440. &  0.0421876 \\
5450. &  0.0564394 \\
5460. &  0.0707405 \\
5470. &  0.0994942 \\
5480. &  0.128835  \\
5490. &  0.159178  \\
\enddata
\end{deluxetable}
\clearpage

\begin{deluxetable}{lc}
\tablewidth{0pt}
\tablecaption{$I$ band response function}
\tablehead{\colhead{Wavelength \AA} & \colhead{Throughput} }
\startdata
6800. &  0 \\
6810. &  0 \\
6820. &  0.000612067 \\
6830. &  0.00220244  \\
6840. &  0.00476994  \\
6850. &  0.00574608  \\
6860. &  0.00672227  \\
6870. &  0.00782015  \\
6880. &  0.0091616   \\
6890. &  0.0107461   \\
\enddata
\end{deluxetable}
\clearpage

\begin{deluxetable}{lc}
\tablewidth{0pt}
\tablecaption{$z$ band response function}
\tablehead{\colhead{Wavelength \AA} & \colhead{Throughput \%}}
\startdata
           8000.  &  0.001212111 \\
           8010.  &  0.001578279 \\
           8020.  &  0.001756357 \\
           8030.  &  0.002116894 \\
           8040.  &  0.002382566 \\ 
           8050.  &  0.002828275 \\
           8060.  &  0.003360764 \\
           8070.  &  0.003797955 \\
           8080.  &  0.004501144 \\
           8090.  &  0.005287436 \\
\enddata
\end{deluxetable}
\clearpage

\begin{deluxetable}{lc}
\tablewidth{0pt}
\tablecaption{$J$ band response function}
\tablehead{\colhead{Wavelength \AA} & \colhead{Throughput \%}}
\startdata
10600.00  & 0 \\
10700.00  & 0.0121698 \\
10800.00  & 0.0854674 \\
10900.00  & 0.478719 \\
11000.00  & 0.531001 \\
11100.00  & 0.54874 \\
11200.00  & 0.00185213 \\
11300.00  & 0.594653  \\
11400.00  & 0.452141  \\
11500.00  & 0.600477  \\
\enddata
\end{deluxetable}
\clearpage

\begin{deluxetable}{lc}
\tablewidth{0pt}
\tablecaption{$H$ band response function}
\tablehead{\colhead{Wavelength \AA} & \colhead{Throughput \%}}
\startdata
13800.00 &  0.000337764 \\
14400.00 &  0.00171936 \\
14500.00 &  0.00537368 \\
14600.00 &  0.00819975 \\
14700.00 &  0.0136237  \\
14800.00 &  0.0368037  \\
14900.00 &  0.0946224  \\
15000.00 &  0.243382   \\
15100.00 &  0.554612   \\
15200.00 &  0.756559   \\
\enddata
\end{deluxetable}
\clearpage

\begin{deluxetable}{lc}
\tablewidth{0pt}
\tablecaption{$K_s$ band response function}
\tablehead{\colhead{Wavelength \AA} & \colhead{Throughput \%}}
\startdata
18500.00 &  0 \\
18600.00 &  0 \\
18700.00 &  1.66052e-06 \\
18800.00 &  0.000684133 \\
18900.00 &  0 \\
19000.00 &  0 \\
19100.00 &  0.00201919 \\
19200.00 &  0 \\
19300.00 &  0.00571882   \\
19400.00 &  0.000896679  \\
\enddata
\end{deluxetable}
\clearpage



\begin{figure}
\epsscale{0.9}
\plotone{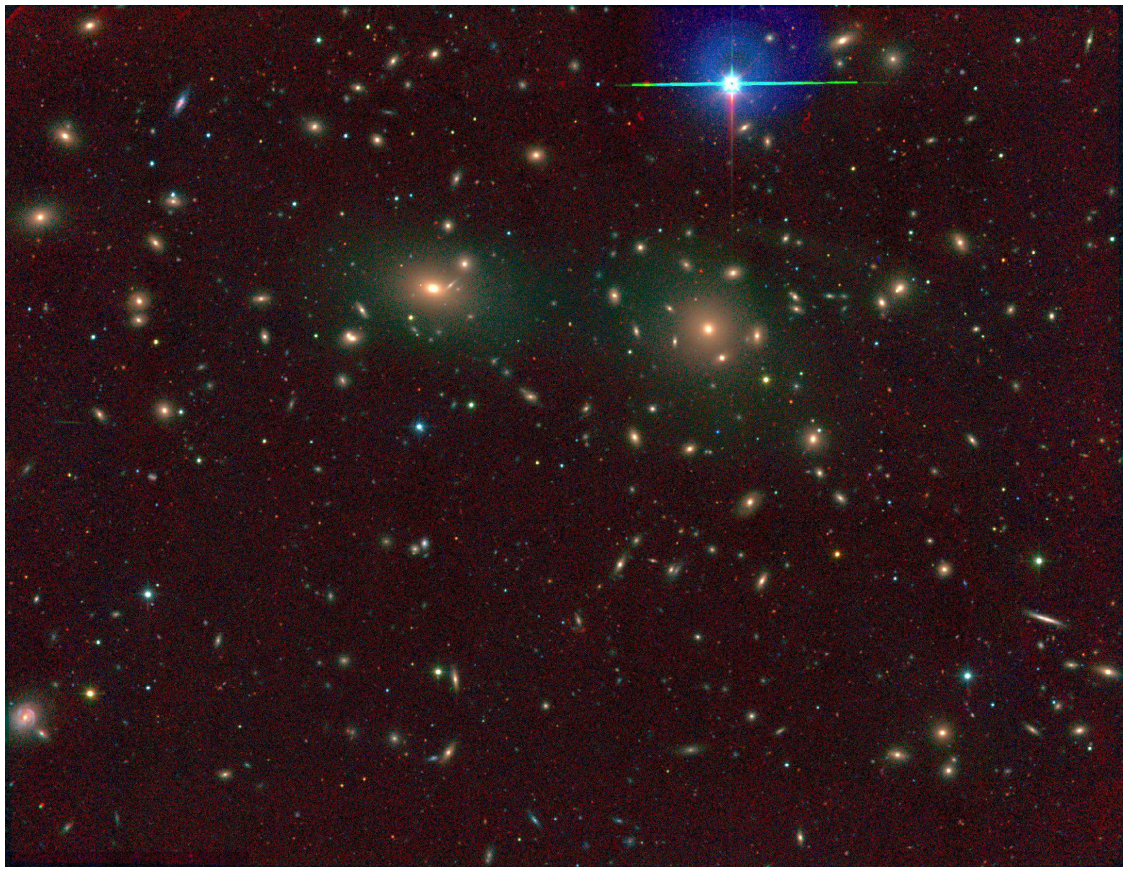}
\caption{A $UVK$ color image of the Coma field used in the present study: here
$U$ is `blue', $V$ is `green' and $K$ is `red'. The image is centered on approximately 
12:59:52.8 +27:55:00 (J2000) and subtends $29.2'$ (RA) $\times 22.5'$ (dec).  North is up, 
and east is to the left. The calibrated fits image mosaics are available at NOAO archives.}
\end{figure}
\clearpage

\begin{figure}
\plotone{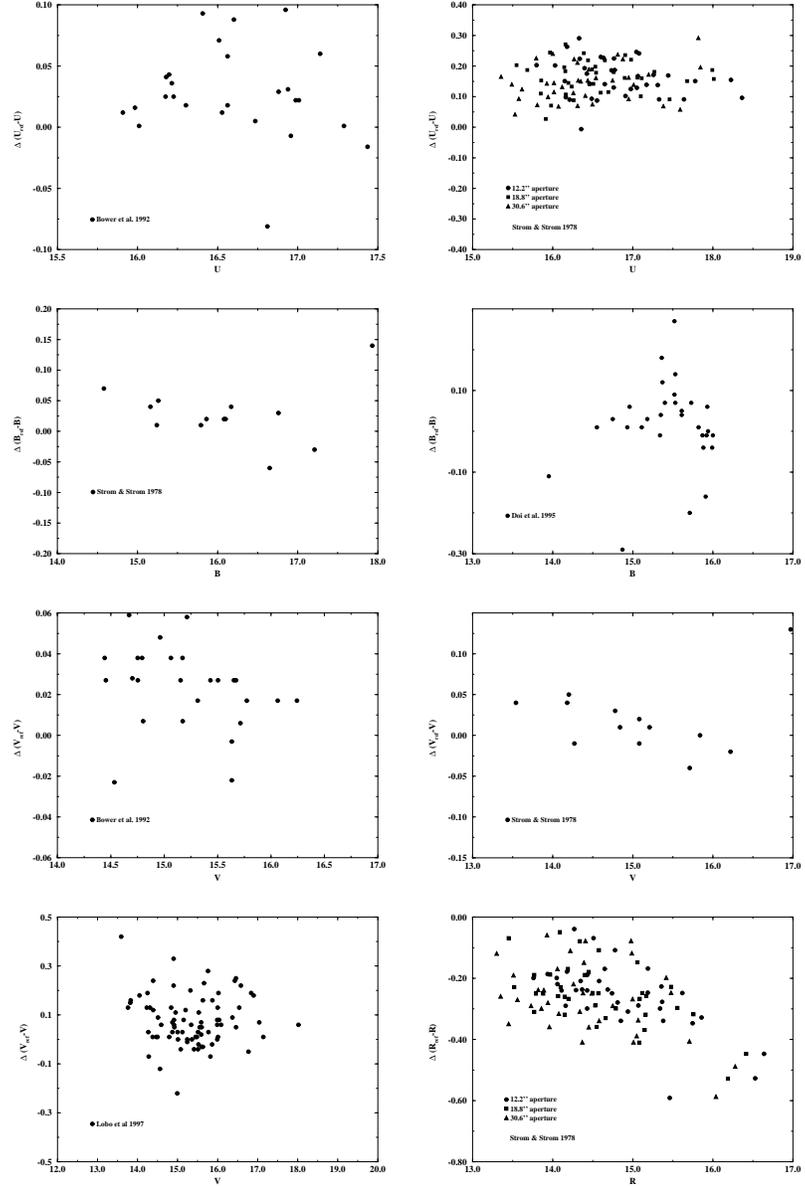}
\caption{Comparison of our photometry with reference photometry (see Table 11)
         The sources for the comparisons are indicated in the Figure legends.}
\end{figure}
\clearpage

\begin{figure}
\figurenum{2}
\plotone{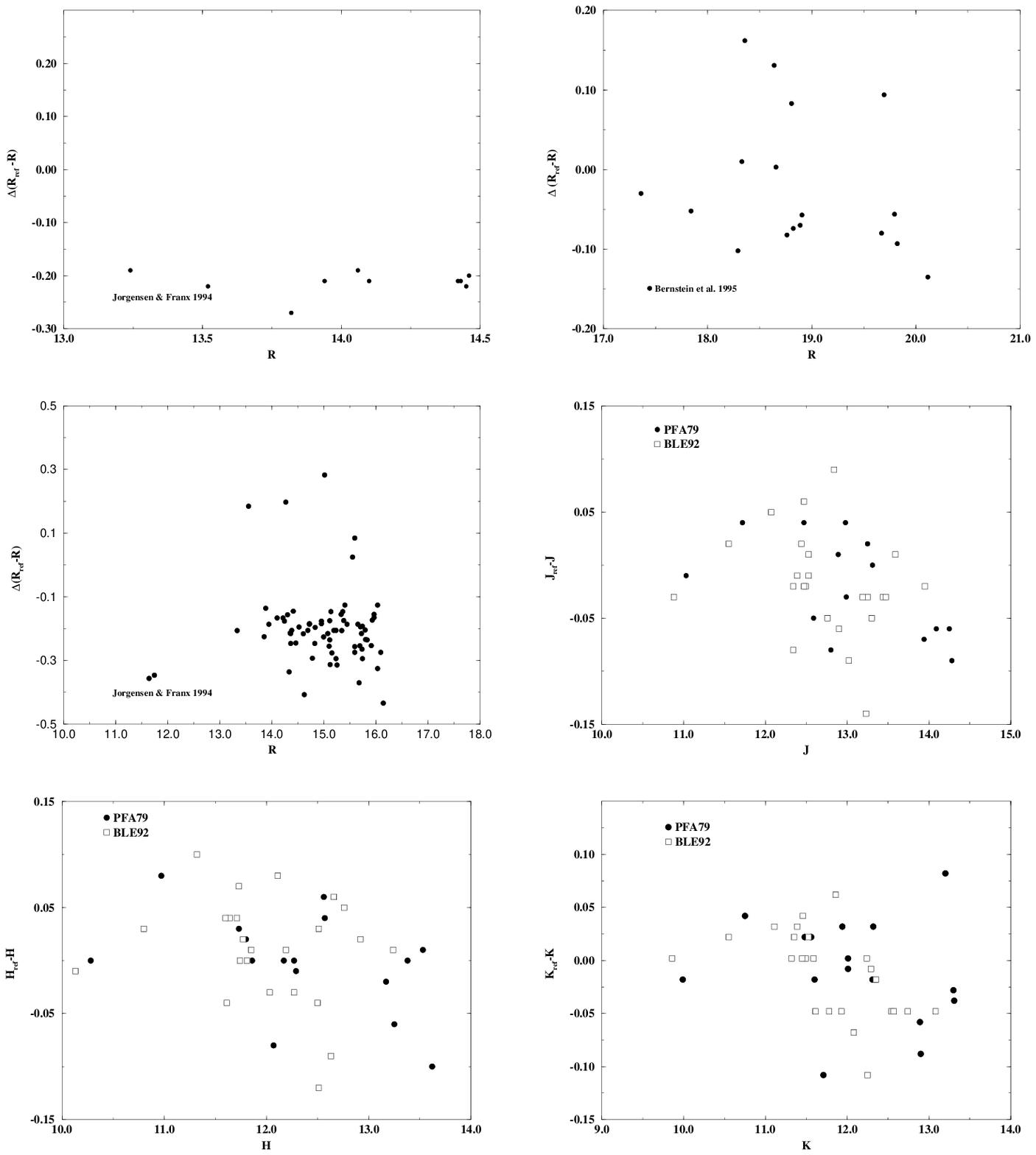}
\caption{Continued}
\end{figure}
\clearpage

\begin{figure}
\plotone{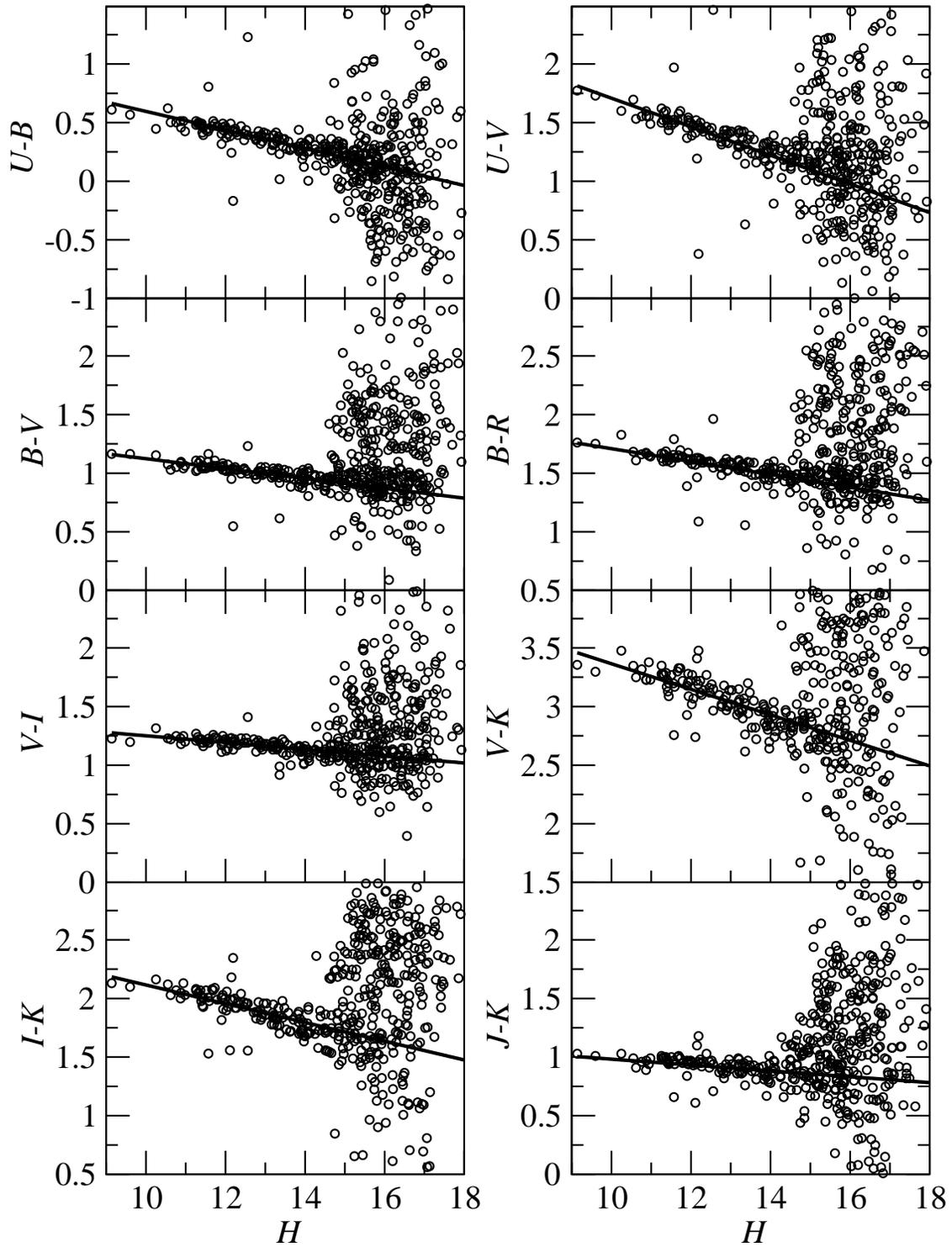}
\caption{Color-magnitude relations for the eight representative colors
tabulated in Table 9. The straight lines show the best fits, whose slopes,
intercepts and scatter are tabulated in Table 12.}
\end{figure}
\clearpage

\begin{figure}
\plotone{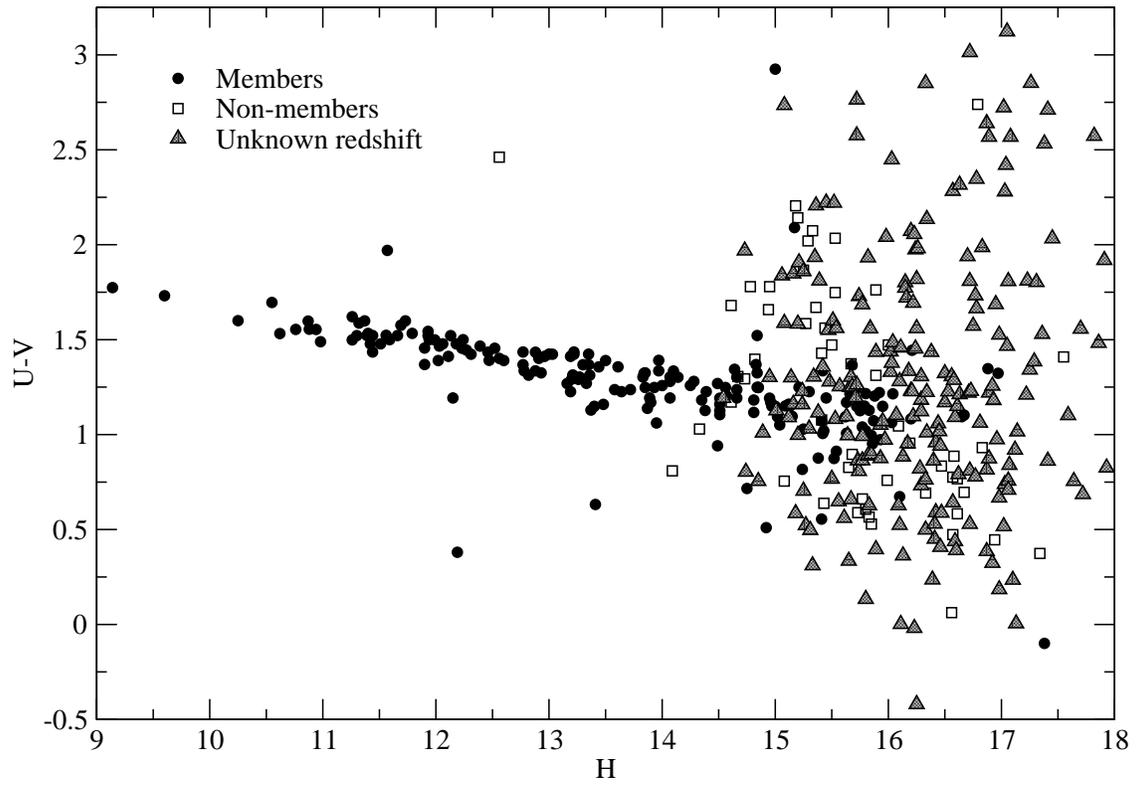}
\caption{Expanded plot for $U-V$ vs. $H$, showing spectroscopic members, non
members and objects with unknown redshifts.}
\end{figure}
\clearpage

\begin{figure}
\plotone{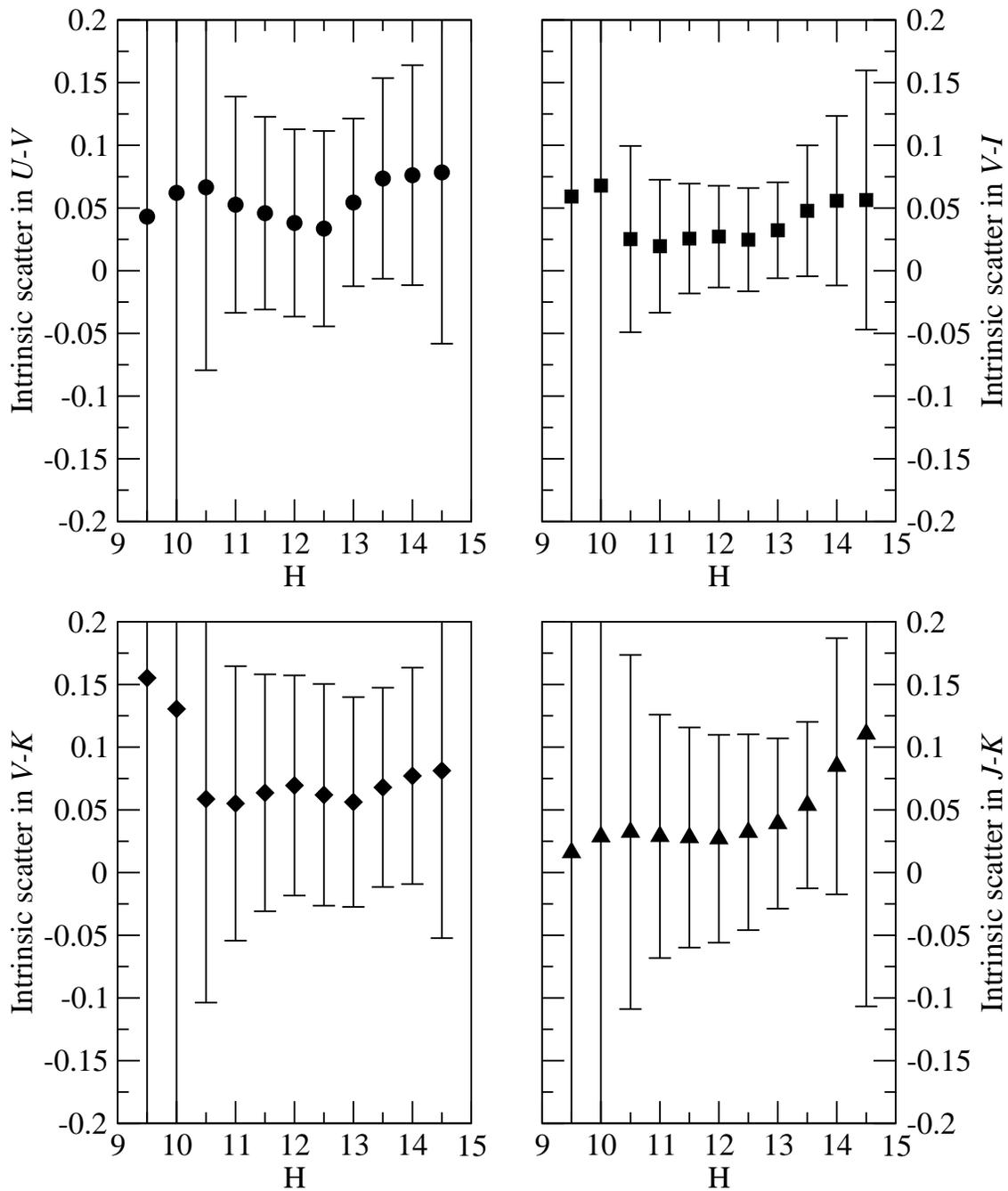}
\caption{Intrinsic scatter about the color-magnitude relation as
a function of $H$-band magnitude, for four colors.}
\end{figure}
\clearpage

\end{document}